\documentclass{article}
\usepackage{amsfonts,amssymb,nips2005e,times,graphics}
\begin{document}
\title{Measuring Shared Information and Coordinated Activity in Neuronal
  Networks}

\author{
Kristina Lisa Klinkner\\
{\small Statistics Department}\\
{\small University of Michigan}\\
{\small Ann Arbor, MI 48109}\\
{\small \texttt{kshalizi@umich.edu}}\\
\And
Cosma Rohilla Shalizi\\
{\small Statistics Department}\\
{\small Carnegie Mellon University}\\
{\small Pittsburgh, PA 15213}\\
{\small \texttt{cshalizi@stat.cmu.edu}}
\And
Marcelo F. Camperi\\
{\small Physics Department}\\
{\small University of San Francisco}\\
{\small San Francisco, CA  94118}\\
{\small \texttt{camperi@usfca.edu}}
}

\maketitle

\begin{abstract} Most nervous systems encode information about stimuli
in the responding activity of large neuronal networks. This activity
often manifests itself as dynamically coordinated sequences of action
potentials.  Since multiple electrode recordings are now a standard
tool in neuroscience research, it is important to have a measure of
such network-wide behavioral coordination and information sharing,
applicable to multiple neural spike train data.  We propose a new
statistic, {\em informational coherence}, which measures how much
better one unit can be predicted by knowing the dynamical state of
another. We argue informational coherence is a measure of association
and shared information which is superior to traditional pairwise
measures of synchronization and correlation.  To find the dynamical
states, we use a recently-introduced algorithm which reconstructs
effective state spaces from stochastic time series.  We then extend
the pairwise measure to a multivariate analysis of the network by
estimating the network multi-information.  We illustrate our method
by testing it on a detailed model of the transition from gamma to beta
rhythms.

\end{abstract}

Much of the most important information in neural systems is shared over
multiple neurons or cortical areas, in such forms as population codes and
distributed representations \cite{Abbott-Sejnowski-neural-codes}.  On
behavioral time scales, neural information is stored in temporal patterns of
activity as opposed to static markers; therefore, as information is shared
between neurons or brain regions, it is physically instantiated as coordination
between entire sequences of neural spikes.  Furthermore, neural systems and
regions of the brain often require coordinated neural activity to perform
important functions; acting in concert requires multiple neurons or cortical
areas to share information \cite{Brown-Kass-Mitra-multiple-spike-train-data}.
Thus, if we want to measure the dynamic network-wide behavior of neurons and
test hypotheses about them, we need reliable, practical methods to detect and
quantify behavioral coordination and the associated information sharing across
multiple neural units. These would be especially useful in testing ideas about
how particular forms of coordination relate to distributed coding (e.g., that
of \cite{Ballard-Zhang-Rao-distributed-synchrony}).  Current techniques to
analyze relations among spike trains handle only pairs of neurons, so we
further need a method which is extendible to analyze the coordination in the
network, system, or region as a whole.  Here we propose a new measure of
behavioral coordination and information sharing, {\em informational coherence},
based on the notion of dynamical state.

Section \ref{sec:synch-measures} argues that coordinated behavior in neural
systems is often not captured by existing measures of synchronization or
correlation, and that something sensitive to nonlinear, stochastic, predictive
relationships is needed.  Section \ref{sec:states-and-IC} defines informational
coherence as the (normalized) mutual information between the dynamical states
of two systems and explains how looking at the states, rather than just
observables, fulfills the needs laid out in Section \ref{sec:synch-measures}.
Since we rarely know the right states {\em a prori}, Section
\ref{sec:reconstruction} briefly describes how we reconstruct effective state
spaces from data. Section \ref{sec:estimating-states-and-IC} gives some details
about how we calculate the informational coherence and approximate the
global information stored in the network.  Section \ref{sec:example} applies
our method to a model system (a biophysically detailed conductance-based model)
comparing our results to those of more familiar second-order statistics.  In
the interest of space, we omit proofs and a full discussion of the existing
literature, giving only minimal references here; proofs and references will
appear in a longer paper now in preparation.

\section{Synchrony or Coherence?}
\label{sec:synch-measures}

Most hypotheses which involve the idea that information sharing is reflected in
coordinated activity across neural units invoke a very specific notion of
coordinated activity, namely strict synchrony: the units should be doing
exactly the same thing (e.g., spiking) at exactly the same time.  Investigators
then measure coordination by measuring how close the units come to being
strictly synchronized (e.g., variance in spike times).

From an informational point of view, there is no reason to favor strict
synchrony over other kinds of coordination.  One neuron consistently spiking 50
ms after another is just as informative a relationship as two simultaneously
spiking, but such stable phase relations are missed by strict-synchrony
approaches.  Indeed, whatever the exact nature of the neural code, it uses
temporally extended patterns of activity, and so information sharing should be
reflected in coordination of those patterns, rather than just the instantaneous
activity.

There are three common ways of going beyond strict synchrony: cross-correlation
and related second-order statistics, mutual information, and topological
generalized synchrony.

The cross-correlation function (the normalized covariance function;
this includes, for present purposes, the joint peristimulus time
histogram
\cite{Brown-Kass-Mitra-multiple-spike-train-data}), is one of the most
widespread measures of synchronization. It can be efficiently
calculated from observable series; it handles statistical as well as
deterministic relationships between processes; by incorporating
variable lags, it reduces the problem of phase locking.  Fourier
transformation of the covariance function $\gamma_{XY}(h)$ yields the
cross-spectrum $F_{XY}(\nu)$, which in turn gives the spectral
coherence $c_{XY}(\nu) = F^2_{XY}(\nu)/F_X(\nu)F_Y(\nu)$, a normalized
correlation between the Fourier components of $X$ and $Y$.  Integrated
over frequencies, the spectral coherence measures, essentially, the
degree of linear cross-predictability of the two series.
(\cite{Brillinger-Villa-assessing-connections} applies spectral
coherence to coordinated neural activity.)  However, such second-order
statistics {\em only} handle linear relationships.  Since neural
processes are known to be strongly nonlinear, there is little reason
to think these statistics adequately measure coordination and
synchrony in neural systems.

Mutual information is attractive because it handles both nonlinear and
stochastic relationships and has a very natural and appealing interpretation.
Unfortunately, it often seems to fail in practice, being disappointingly small
even between signals which are known to be tightly coupled
\cite{Quian-Quiroga-Grassberger-synch}. The major reason is that the neural
codes use distinct patterns of activity over time, rather than many different
instantaneous actions, and the usual approach misses these extended patterns.
Consider two neurons, one of which drives the other to spike 50 ms after it
does, the driving neuron spiking once every 500 ms.  These are very tightly
coordinated, but whether the first neuron spiked at time $t$ conveys little
information about what the second neuron is doing at $t$ --- it's not spiking,
but it's not spiking most of the time anyway.  Mutual information calculated
from the direct observations conflates the ``no spike'' of the second neuron
preparing to fire with its just-sitting-around ``no spike''.  Here, mutual
information could find the coordination if we used a 50 ms lag, but that won't
work in general.  Take two rate-coding neurons with base-line firing rates of 1
Hz, and suppose that a stimulus excites one to 10 Hz and suppresses the other
to 0.1 Hz.  The spiking rates thus share a lot of information, but whether the
one neuron spiked at $t$ is uninformative about what the other neuron did then,
and lagging won't help.

Generalized synchrony is based on the idea of establishing relationships
between the states of the various units.  ``State'' here is taken in the sense
of physics, dynamics and control theory: the state at time $t$ is a variable
which fixes the distribution of observables at all times $\geq t$, rendering
the past of the system irrelevant \cite{Streater-statistical-dynamics}.
Knowing the state allows us to predict, as well as possible, how the system
will evolve, and how it will respond to external forces
\cite{predictive-representations-of-state}.  Two coupled systems are said to
exhibit generalized synchrony if the state of one system is given by a mapping
from the state of the other.  Applications to data employ state-space
reconstruction \cite{Kantz-Schreiber}: if the state $x \in \mathcal{X}$ evolves
according to smooth, $d$-dimensional deterministic dynamics, and we observe a
generic function $y = f(x)$, then the space $\mathcal{Y}$ of time-delay vectors
$\left[ y(t), y(t-\tau), ...  y(t-(k-1)\tau) \right]$ is diffeomorphic to
$\mathcal{X}$ if $k > 2d$, for generic choices of lag $\tau$.  The various
versions of generalized synchrony differ on how, precisely, to quantify the
mappings between reconstructed state spaces, but they all appear to be
empirically equivalent to one another and to notions of phase synchronization
based on Hilbert transforms \cite{Quian-Quiroga-Grassberger-synch}. Thus all of
these measures accommodate nonlinear relationships, and are potentially very
flexible.  Unfortunately, there is essentially no reason to believe that neural
systems have deterministic dynamics at experimentally-accessible levels of
detail, much less that there are deterministic relationships among such states
for different units.

What we want, then, but none of these alternatives provides, is a quantity
which measures predictive relationships among states, but allows those
relationships to be nonlinear and stochastic.  The next section introduces just
such a measure, which we call ``informational coherence''.

\section{States and Informational Coherence}
\label{sec:states-and-IC}

There are alternatives to calculating the ``surface'' mutual information
between the sequences of observations themselves (which, as described, fails to
capture coordination).  If we know that the units are phase oscillators, or
rate coders, we can estimate their instantaneous phase or rate and, by
calculating the mutual information between those variables, see how coordinated
the units' patterns of activity are.  However, phases and rates do not exhaust
the repertoire of neural patterns and a more general, common scheme is
desirable.  The most general notion of ``pattern of activity'' is simply that
of the dynamical state of the system, in the sense mentioned above. We now
formalize this.

Assuming the usual notation for Shannon information \cite{Cover-and-Thomas},
the information content of a state variable $X$ is $H[X]$ and the mutual
information between $X$ and $Y$ is $I[X;Y]$.  As is well-known, $I[X;Y] \leq
\min{H[X],H[Y]}$.  We use this to normalize the mutual state information to a
$0-1$ scale, and this is the {\em informational coherence} (IC).
\begin{eqnarray}
\psi(X,Y) & = & \frac{I[X;Y]}{\min{H[X],H[Y]}}~, ~ ~ \mathrm{with}\ 0/0 = 0~.
\end{eqnarray}

$\psi$ can be interpreted as follows.  $I[X;Y]$ is the Kullback-Leibler
divergence between the joint distribution of $X$ and $Y$, and the product of
their marginal distributions \cite{Cover-and-Thomas}, indicating the error
involved in ignoring the dependence between $X$ and $Y$.  The mutual
information between predictive, dynamical states thus gauges the error involved
in assuming the two systems are independent, i.e., how much predictions could
improve by taking into account the dependence.  Hence it measures the amount of
{\em dynamically-relevant} information shared between the two systems.  $\psi$
simply normalizes this value, and indicates the degree to which two systems
have coordinated {\em patterns} of behavior (cf.\ \cite{Palus-et-al-synch-as-adj-of-info-rates}, although this only uses directly observable quantities).

\subsection{Reconstruction and Estimation of Effective State Spaces}
\label{sec:reconstruction}

As mentioned, the state space of a deterministic dynamical system can be
reconstructed from a sequence of observations.  This is the main tool of
experimental nonlinear dynamics \cite{Kantz-Schreiber}; but the assumption of
determinism is crucial and false, for almost any interesting neural system.
While classical state-space reconstruction won't work on stochastic processes,
such processes {\em do} have state-space representations
\cite{Knight-predictive-view}, and, in the special case of discrete-valued,
discrete-time series, there are ways to reconstruct the state space.

Here we use the CSSR algorithm, introduced in \cite{CSSR-for-UAI} (code
available at \texttt{http://bactra.org/CSSR}).  This produces {\em causal state
  models}, which are stochastic automata capable of statistically-optimal
nonlinear prediction; the state of the machine is a minimal sufficient
statistic for the future of the observable
process\cite{CMPPSS}.\footnote{Causal state models have the same expressive
  power as observable operator models \cite{Jaeger-operator-models} or
  predictive state representations \cite{predictive-representations-of-state},
  and greater power than variable-length Markov models
  \cite{Ron-Singer-Tishby-amnesia,Buhlmann-Wyner}.}  The basic idea is to form
a set of states which should be (1) Markovian, (2) sufficient statistics for
the next observable, and (3) have deterministic transitions (in the
automata-theory sense).  The algorithm begins with a minimal, one-state, IID
model, and checks whether these properties hold, by means of hypothesis tests.
If they fail, the model is modified, generally but not always by adding more
states, and the new model is checked again.  Each state of the model
corresponds to a distinct distribution over future events, i.e., to a
statistical pattern of behavior.  Under mild conditions, which do not involve
prior knowledge of the state space, CSSR converges in probability to the unique
causal state model of the data-generating process \cite{CSSR-for-UAI}.  In
practice, CSSR is quite fast (linear in the data size), and generalizes at
least as well as training hidden Markov models with the EM algorithm and using
cross-validation for selection, the standard heuristic \cite{CSSR-for-UAI}.

One advantage of the causal state approach (which it shares with classical
state-space reconstruction) is that state estimation is greatly simplified.  In
the general case of nonlinear state estimation, it is necessary to know not
just the form of the stochastic dynamics in the state space and the observation
function, but also their precise parametric values and the distribution of
observation and driving noises.  Estimating the state from the observable time
series then becomes a computationally-intensive application of Bayes's Rule
\cite{Ahmed-filtering}.

Due to the way causal states are built as statistics of the data, with
probability 1 there is a finite time, $t$, at which the causal state at time
$t$ is certain.  This is not just with some degree of belief or confidence:
because of the way the states are constructed, it is impossible for the process
to be in any other state at that time.  Once the causal state has been
established, it can be updated recursively, i.e., the causal state at time
$t+1$ is an explicit function of the causal state at time $t$ and the
observation at $t+1$.  The causal state model can be automatically converted,
therefore, into a finite-state transducer which reads in an observation time
series and outputs the corresponding series of states
\cite{Upper-thesis,CMPPSS}.  (Our implementation of CSSR filters its training
data automatically.)  The result is a new time series of states, from which all
non-predictive components have been filtered out.

\subsection{Estimating the Coherence}
\label{sec:estimating-states-and-IC}

Our algorithm for estimating the matrix of informational coherences is as
follows.  For each unit, we reconstruct the causal state model, and filter the
observable time series to produce a series of causal states.  Then, for each
pair of neurons, we construct a joint histogram of the state distribution,
estimate the mutual information between the states, and normalize by the
single-unit state informations.  This gives a symmetric matrix of $\psi$
values.

Even if two systems are independent, their estimated IC will, on
average, be positive, because, while they should have zero mutual
information, the empirical estimate of mutual information is
non-negative.  Thus, the significance of IC values must be assessed
against the null hypothesis of system independence.  The easiest way
to do so is to take the reconstructed state models for the two systems
and run them forward, independently of one another, to generate a
large number of simulated state sequences; from these calculate
values of the IC.  This procedure will approximate the sampling
distribution of the IC under a null model which preserves the dynamics
of each system, but not their interaction.  We can then find
$p$-values as usual. We omit them here to save space.

\subsection{Approximating the Network Multi-Information}
\label{sec:multi-information}

\begin{figure}
\begin{center}$a$\resizebox{0.45\columnwidth}{!}{\includegraphics{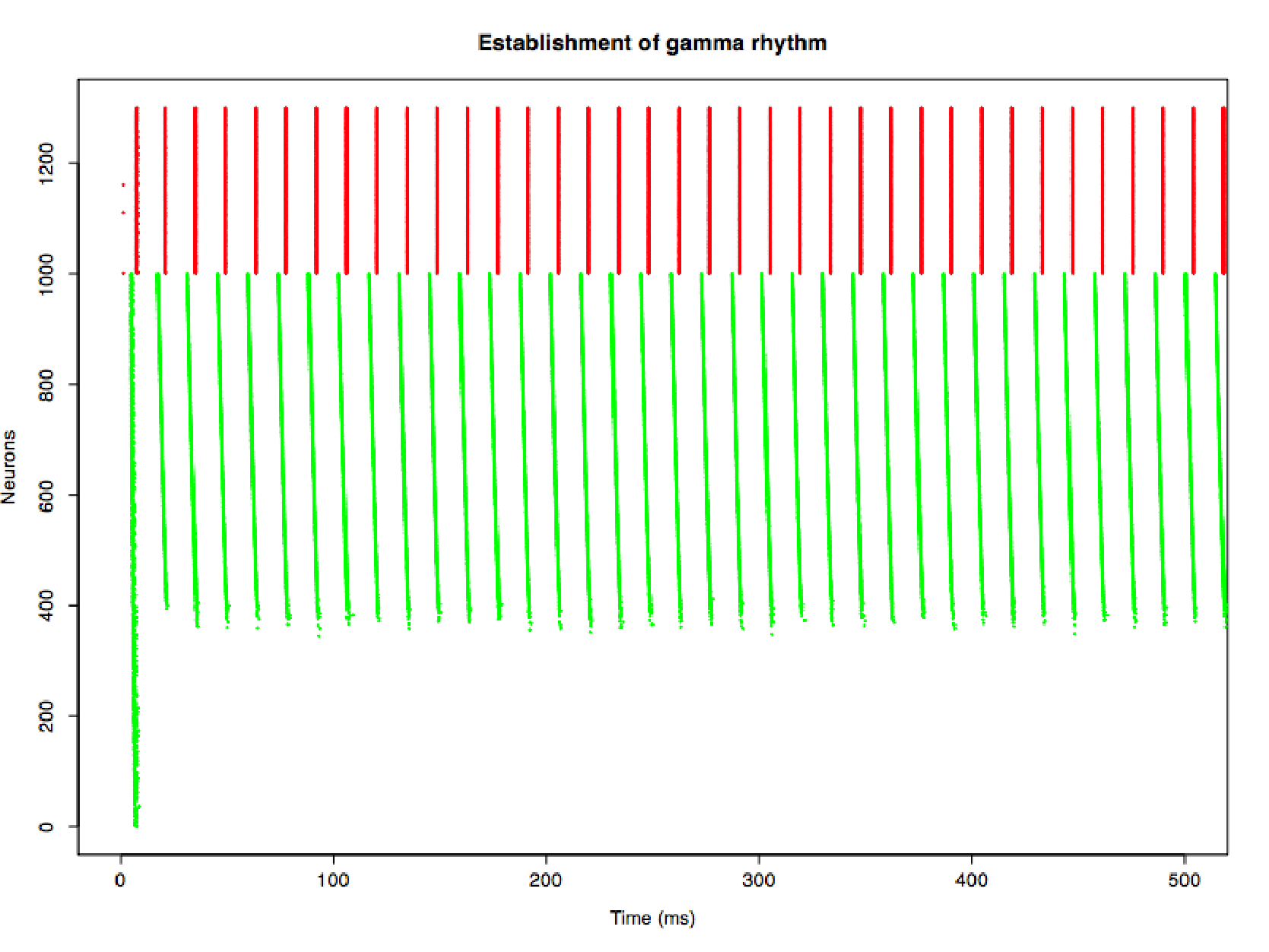}} $b$\resizebox{0.45\columnwidth}{!}{\includegraphics{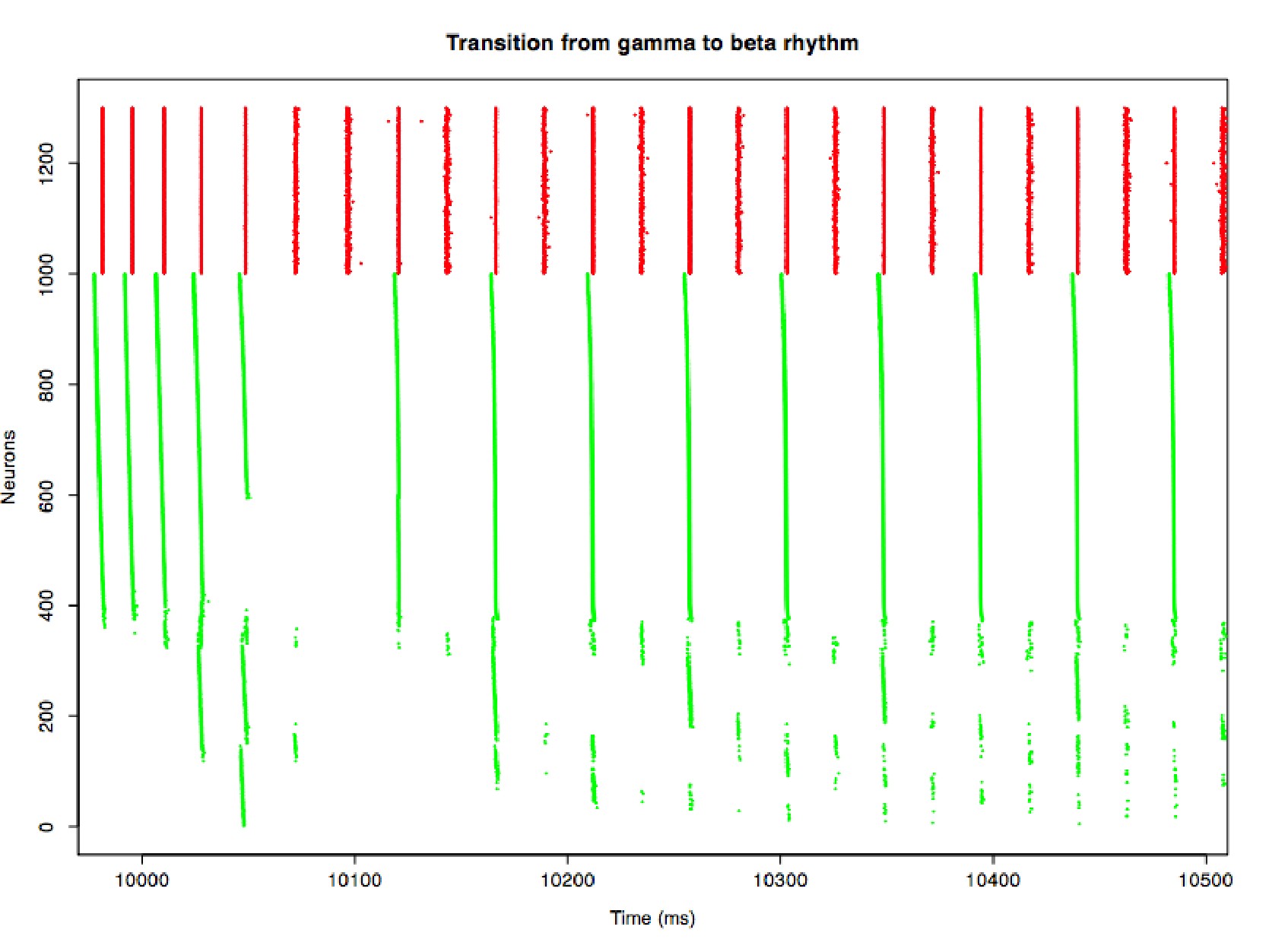}}\end{center}
\caption{{\small Rastergrams of neuronal spike-times in the network.
    Excitatory, pyramidal neurons (numbers 1 to 1000) are shown in green,
    inhibitory interneurons (numbers 1001 to 1300) in red.  During the first 10
    seconds ($a$), the current connections among the pyramidal cells are
    suppressed and a gamma rhythm emerges (left).  At $t=10\mathrm{s}$, those
    connections become active, leading to a beta rhythm ($b$, right).}}
\label{fig:rastergram}
\end{figure}

There is broad agreement \cite{Brown-Kass-Mitra-multiple-spike-train-data} that
analyses of networks should not just be an analysis of pairs of neurons,
averaged over pairs. Ideally, an analysis of information sharing in a network
would look at the over-all structure of statistical dependence between the
various units, reflected in the complete joint probability distribution $P$ of
the states.  This would then allow us, for instance, to calculate the $n$-fold
multi-information, $I[X_1, X_2, \ldots X_n] \equiv D(P||Q)$, the
Kullback-Leibler divergence between the joint distribution $P$ and the product
of marginal distributions $Q$, analogous to the pairwise mutual information
\cite{Schneidman-et-al-network-information}.  Calculated over the predictive
states, the multi-information would give the total amount of shared dynamical
information in the system.  Just as we normalized the mutual information
$I[X_1, X_2]$ by its maximum possible value, $\min{H[X_1], H[X_2]}$, we
normalize the multi-information by its maximum, which is the smallest sum of $n-1$ marginal entropies:
\[
I[X_1; X_2; \ldots X_n] \leq \min_{k}{\sum_{i\neq k}{H[X_n]}}
\]
Unfortunately, $P$ is a distribution over a very high dimensional space and so, hard to estimate well without strong parametric constraints.  We thus consider
approximations.

The lowest-order approximation treats all the units as independent; this is the
distribution $Q$.  One step up are tree distributions, where the global
distribution is a function of the joint distributions of pairs of units.  Not
every pair of units needs to enter into such a distribution, though every unit
must be part of some pair.  Graphically, a tree distribution corresponds to a
spanning tree, with edges linking units whose interactions enter into the
global probability, and conversely spanning trees determine tree distributions.
Writing $E_T$ for the set of pairs $(i, j)$ and abbreviating $X_1 = x_1, X_2 =
x_2, \ldots X_n = x_n$ by $\mathbf{X} = \mathbf{x}$, one has
\begin{equation}
T(\mathbf{X} = \mathbf{x}) = \prod_{(i,j) \in E_T}{\frac{T(X_i = x_i, X_j = x_j)}{T(X_i = x_i)T(X_j = x_j)}}\prod_{i = 1}^{n}{T(X_i = x_i)}
\label{eqn:chow-liu-prob-dist}
\end{equation}
where the marginal distributions $T(X_i)$ and the pair distributions $T(X_i,
X_j)$ are estimated by the empirical marginal and pair distributions.

We must now pick edges $E_T$ so that $T$ best approximates the true global
distribution $P$.  A natural approach is to minimize $D(P||T)$, the divergence
between $P$ and its tree approximation.  Chow and Liu \cite{Chow-Liu-trees}
showed that the maximum-weight spanning tree gives the divergence-minimizing
distribution, taking an edge's weight to be the mutual information between the
variables it links.

There are three advantages to using the Chow-Liu approximation.  (1) Estimating
$T$ from empirical probabilities gives a consistent maximum likelihood
estimator of the ideal Chow-Liu tree \cite{Chow-Liu-trees}, with reasonable
rates of convergence, so $T$ can be reliably known even if $P$ cannot.  (2)
There are efficient algorithms for constructing maximum-weight spanning trees,
such as Prim's algorithm \cite[sec.\ 23.2]{Algorithms-Calder}, which runs in
time $O(n^2 + n\log{n})$.  Thus, the approximation is computationally
tractable.  (3) The KL divergence of the Chow-Liu distribution from $Q$ gives a
lower bound on the network multi-information; that bound is just the sum of
the mutual informations along the edges in the tree:
\begin{equation}
I[X_1; X_2; \ldots X_n] \geq D(T||Q) = \sum_{(i,j) \in E_T}{I[X_i;X_j]}
\label{eqn:chow-liu-approx-to-multi-info}
\end{equation}
Even if we knew $P$ exactly, Eq.\ \ref{eqn:chow-liu-approx-to-multi-info} would
be useful as an alternative to calculating $D(P||Q)$ directly, evaluating
$\log{P(\mathbf{x})/Q(\mathbf{x})}$ for all the exponentially-many
configurations $\mathbf{x}$.

It is natural to seek higher-order approximations to $P$, e.g., using three-way
interactions not decomposable into pairwise interactions
\cite{Amari-hierarchical-info-geo,Schneidman-et-al-network-information}.  But
it is hard to do so effectively, because finding the optimal approximation to
$P$ when such interactions are allowed is NP
\cite{Kirshner-Smyth-Robertson-on-Chow-Liu-trees-TR}, and analytical formulas
like Eq.\ \ref{eqn:chow-liu-approx-to-multi-info} generally do not exist
\cite{Schneidman-et-al-network-information}.  We therefore confine ourselves to
the Chow-Liu approximation here.

\begin{figure}[t]
\begin{center}{\em a}\resizebox{0.45\columnwidth}{!}{\includegraphics{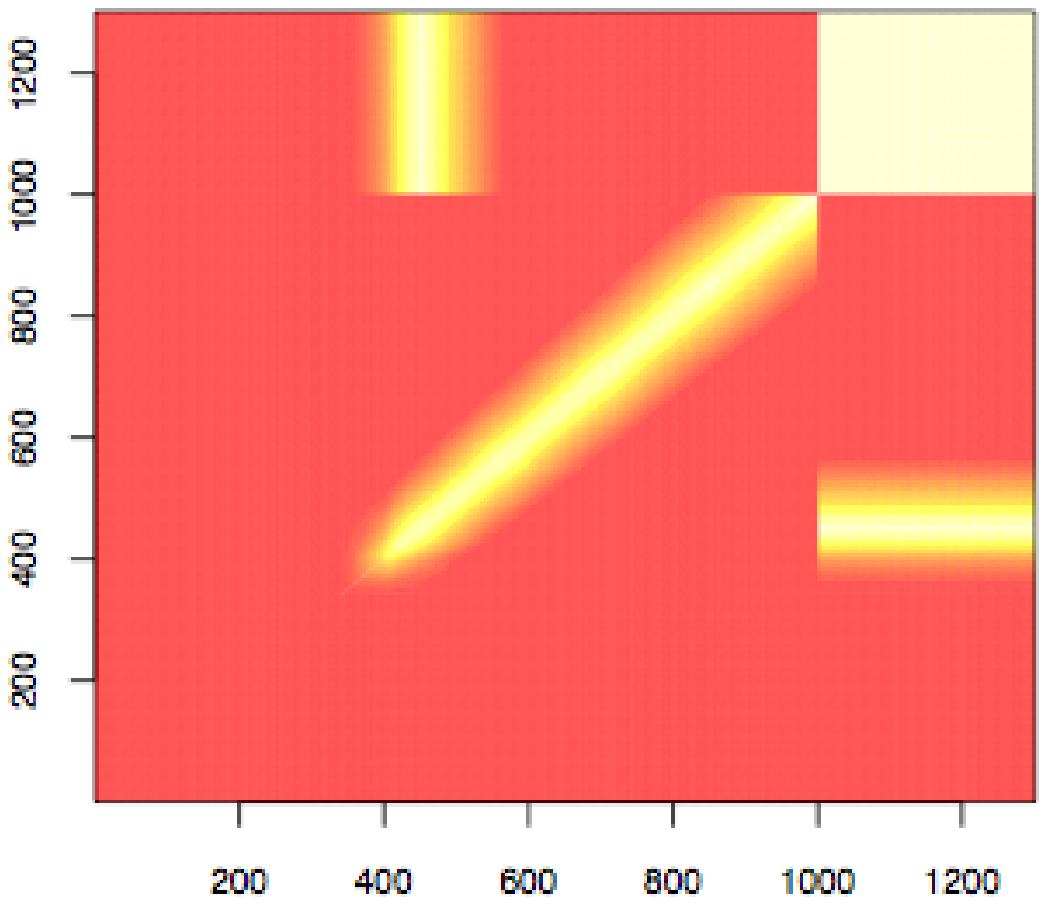}}
{\em b}\resizebox{0.45\columnwidth}{!}{\includegraphics{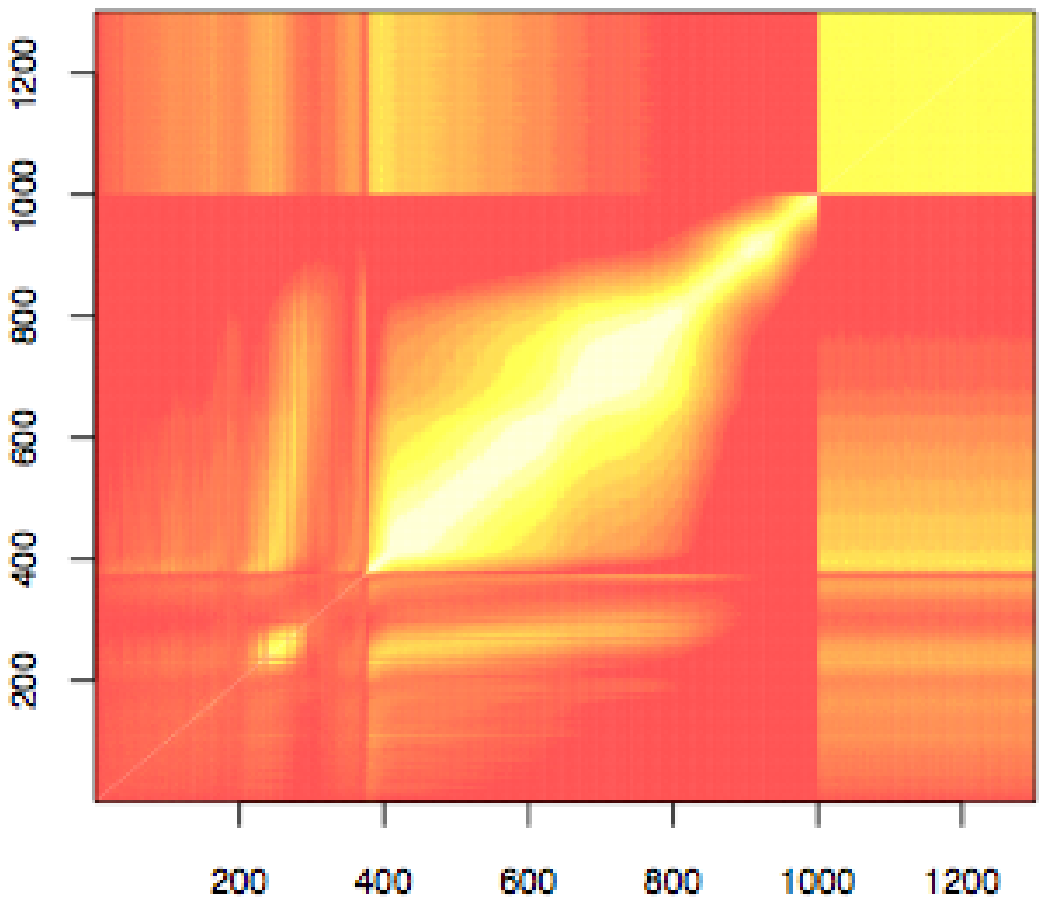}}\\
{\em c}\resizebox{0.45\columnwidth}{!}{\includegraphics{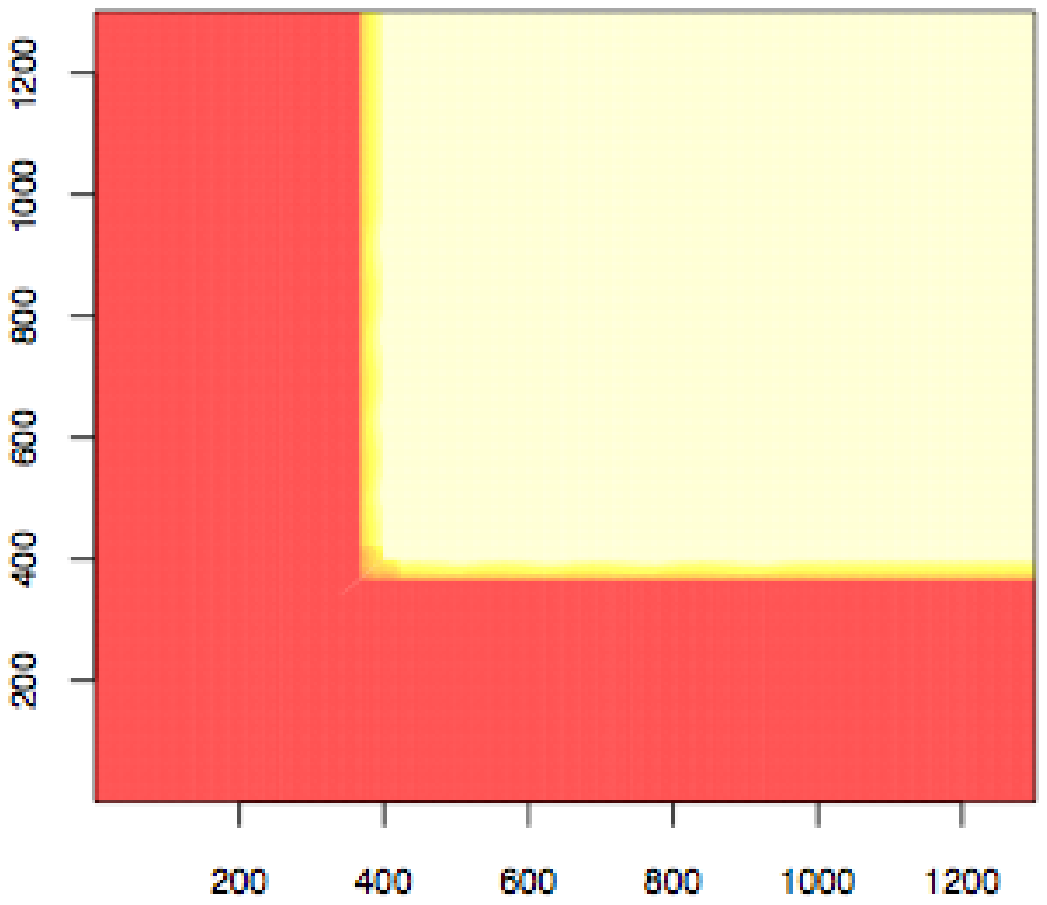}}
{\em d}\resizebox{0.45\columnwidth}{!}{\includegraphics{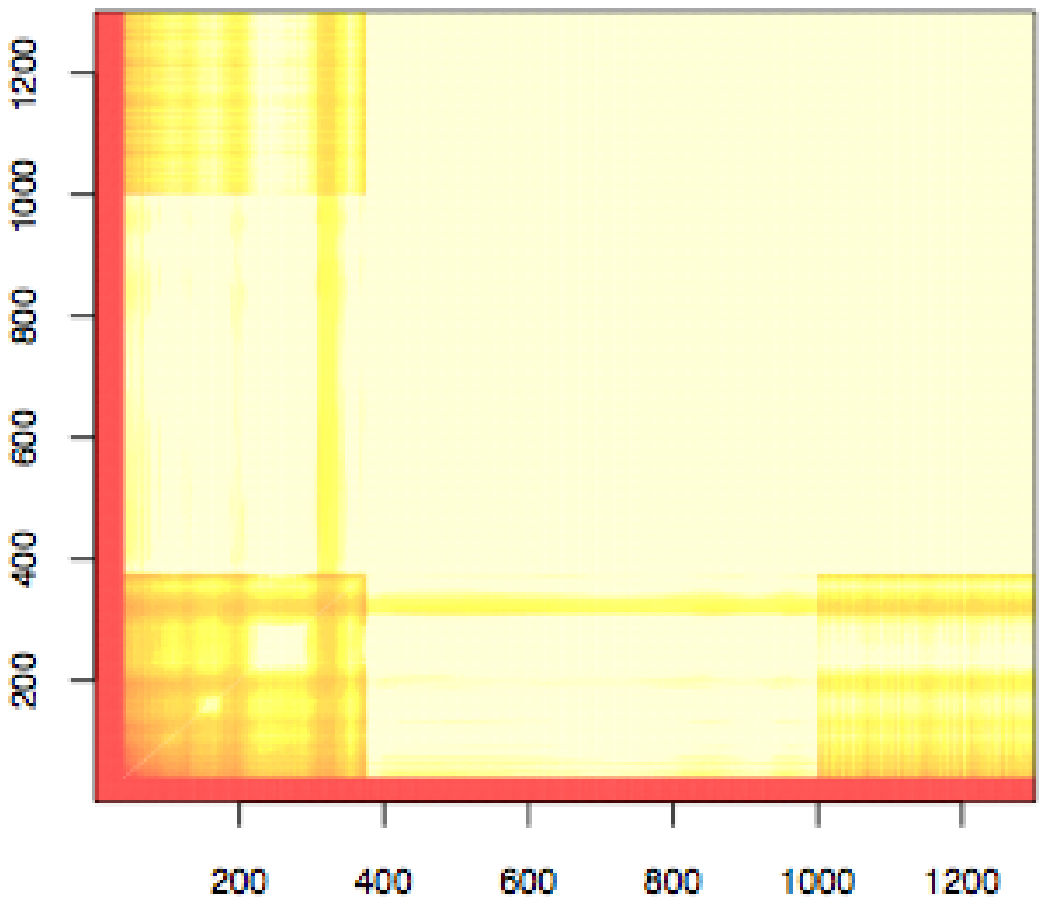}}
\end{center}
\caption{{\small Heat-maps of coordination for the network, as measured by
    zero-lag cross-correlation (top row) and informational coherence (bottom),
    contrasting the gamma rhythm (left column) with the beta (right).  Colors
    run from red (no coordination) through yellow to pale cream (maximum).}}
\label{fig:heat-maps-first-stage}
\end{figure}

\section{Example: A Model of Gamma and Beta Rhythms}
\label{sec:example}

We use simulated data as a test case, instead of empirical multiple electrode
recordings, which allows us to try the method on a system of over 1000
neurons and compare the measure against expected results.  The model,
taken from
\cite{New-roles-for-the-gamma-rhythm}, was originally designed to study
episodes of gamma (30--80Hz) and beta (12--30Hz) oscillations in the
mammalian nervous system, which often occur successively with a
spontaneous transition between them. More concretely, the rhythms
studied were those displayed by {\it in vitro} hippocampal (CA1) slice
preparations and by {\it in vivo} neocortical EEGs.

The model contains two neuron populations: excitatory (AMPA) pyramidal neurons
and inhibitory (GABA$_{\mbox{\footnotesize{A}}}$) interneurons, defined by
conductance-based Hodgkin-Huxley-style equations. Simulations were carried out
in a network of 1000 pyramidal cells and 300 interneurons.  Each cell was
modeled as a one-compartment neuron with all-to-all coupling, endowed with the
basic sodium and potassium spiking currents, an external applied current, and
some Gaussian input noise.

The first 10 seconds of the simulation correspond to the gamma rhythm,
in which only a group of neurons is made to spike via a linearly
increasing applied current.  The beta rhythm (subsequent 10 seconds)
is obtained by activating pyramidal-pyramidal recurrent connections
(potentiated by Hebbian preprocessing as a result of synchrony during
the gamma rhythm) and a slow outward after-hyper-polarization (AHP)
current (the M-current), suppressed during gamma due to the
metabotropic activation used in the generation of the rhythm.  During
the beta rhythm, pyramidal cells, silent during gamma rhythm, fire on
a subset of interneurons cycles (Fig.\ \ref{fig:rastergram}).

Fig.\ \ref{fig:heat-maps-first-stage} compares zero-lag cross-correlation, a
second-order method of quantifying coordination, with the informational
coherence calculated from the reconstructed states.  (In this simulation, we
could have calculated the actual states of the model neurons directly, rather
than reconstructing them, but for purposes of testing our method we did not.)
Cross-correlation finds some of the relationships visible in
Fig.\ \ref{fig:rastergram}, but is confused by, for instance, the phase shifts
between pyramidal cells.  (Surface mutual information, not shown, gives similar
results.)  Informational coherence, however, has no trouble recognizing the two
populations as effectively coordinated blocks.  The presence of dynamical
noise, problematic for ordinary state reconstruction, is not an issue.  The
average IC is $0.411$ (or $0.797$ if the inactive, low-numbered neurons are
excluded).  The tree estimate of the global informational multi-information is
$3243.7$ bits, with a global coherence of $0.777$.  The right half of
Fig.\ \ref{fig:heat-maps-first-stage} repeats this analysis for the beta
rhythm; in this stage, the average IC is $0.614$, and the tree estimate of the
global multi-information is $7377.7$ bits, though the estimated global
coherence falls very slightly to $0.742$.  This is because low-numbered neurons
which were quiescent before are now active, contributing to the global
information, but the over-all pattern is somewhat weaker and more noisy (as can
be seen from Fig.\ \ref{fig:rastergram}$b$.)  So, as expected, the total
information content is higher, but the overall coordination across the network
is lower.

\section{Conclusion}

Informational coherence provides a measure of neural information sharing and
coordinated activity which accommodates nonlinear, stochastic relationships
between extended patterns of spiking.  It is robust to dynamical noise and
leads to a genuinely multivariate measure of global coordination across
networks or regions.  Applied to data from multi-electrode recordings, it
should be a valuable tool in evaluating hypotheses about distributed neural
representation and function.

\subsubsection*{Acknowledgments}

\small{Thanks to R. Haslinger, E. Ionides and S. Page; and for support to the
  Santa Fe Institute (under grants from Intel, the NSF and the MacArthur
  Foundation, and DARPA agreement F30602-00-2-0583), the Clare Booth Luce
  Foundation (KLK) and the James S. McDonnell Foundation (CRS).}

\setlength{\itemsep}{-1mm}

\end{document}